\documentclass[a4paper,journal]{IEEEtran} 
\usepackage{graphicx}
\usepackage{dcolumn}
\usepackage{bm}
\usepackage[utf8]{inputenc}
\usepackage[T1]{fontenc}
\usepackage{mathptmx}
\usepackage{etoolbox}
\usepackage[usenames,dvipsnames]{xcolor} 
\makeatletter
\def\@email#1#2{%
 \endgroup
 \patchcmd{\titleblock@produce}
  {\frontmatter@RRAPformat}
  {\frontmatter@RRAPformat{\produce@RRAP{*#1\href{mailto:#2}{#2}}}\frontmatter@RRAPformat}
  {}{}
}%
\makeatother
\usepackage[usenames,dvipsnames]{xcolor} 

\newcommand{\jon}[1]{\textcolor{black}{#1}}

\newcommand{\jonE}[1]{\textcolor{black}{#1}}
\newcommand{\gn}[1]{\textcolor{black}{#1}}
\newcommand{\dd}[1]{\textcolor{black}{#1}}

\usepackage{amsmath,amsfonts}
\usepackage{algorithmic}
\usepackage{algorithm}
\usepackage{array}
\usepackage[caption=false,font=normalsize,labelfont=sf,textfont=sf]{subfig}
\usepackage{textcomp}
\usepackage{stfloats}
\usepackage{url}
\usepackage{verbatim}
\usepackage{graphicx}
\usepackage{cite}
\usepackage{hyperref}

\IEEEoverridecommandlockouts

\begin{document}


\title{Method for efficient large-scale cryogenic characterization of CMOS technologies}

\author{Jonathan Eastoe, Grayson M. Noah, Debargha Dutta, Alessandro Rossi, Jonathan D. Fletcher, Alberto Gomez-Saiz%

\thanks{Jonathan Eastoe and Jonathan D. Fletcher are with the National Physical Laboratory, Teddington, UK}
\thanks{Alessandro Rossi (\href{mailto:alessandro.rossi@npl.co.uk}{alessandro.rossi@npl.co.uk}) is with the National Physical Laboratory, Teddington, UK and also Department of Physics, SUPA, University of Strathclyde, Glasgow, UK}
\thanks{Grayson M. Noah (\href{mailto:grayson@quantummotion.tech}{grayson@quantummotion.tech}), Debargha Dutta, and Alberto Gomez-Saiz are with Quantum Motion, 9 Sterling Way, London N7 9HJ, UK}%
} 



\markboth{IEEE Draft}{}

\date{\today}

\maketitle
\begin{abstract}
Semiconductor integrated circuits operated at cryogenic temperature will play an essential role in quantum computing architectures. These can offer equivalent or superior performance to their room-temperature counterparts while enabling a scaling up of the total number of qubits under control. Silicon integrated circuits can be operated at a temperature stage of a cryogenic system where cooling power is sufficient ($\sim$3.5+~K) to allow for analog signal chain components (e.g. amplifiers and mixers), local signal synthesis, signal digitization, and control logic. A critical stage in cryo-electronics development is the characterization of individual transistor devices in a particular technology node at cryogenic temperatures. This data enables the creation of a process design kit (PDK) to model devices and simulate integrated circuits operating well below the minimum standard temperature ranges covered by foundry-released models (e.g. -55 °C). Here, an efficient approach to the characterization of large numbers of components at cryogenic temperature is reported. We developed a system to perform DC measurements with Kelvin sense of individual transistors at 4.2~K using integrated on-die multiplexers, enabling bulk characterization of thousands of devices with no physical change to the measurement setup.
\end{abstract}


\section{Introduction}

Quantum computing devices based in semiconductor systems offer many possible advantages of ready scalability by harnessing existing fabrication technology~\cite{GonzalezZalba2021}. Application-specific integrated circuits (ASICs) fabricated using established foundry processes can be operated at low temperature to support the scalable control of the large number of qubits necessary for error correction~\cite{devitt2013quantum}. Integrated multiplexing and signal processing reduces the number of connecting lines between cryogenic stages and room-temperature instruments and reduces the complexity of the room-temperature readout instrumentation required \cite{pauka2021cryogenic}. Significant progress has been made in the development of quantum-orientated CMOS-based cryogenic control systems~\cite{CharbonIEDM2016,CharbonISSC2017,Patra2018, 9772841} which can be utilised in a variety of quantum device platforms including superconducting qubits \cite{AcharyaNE2023, 9731538} and trapped ions \cite{MehtaAPL2014}.

At the most fundamental level, confident cryogenic integrated circuit design begins with test and validation of the elementary transistor components in the technology node under realistic cryogenic operation conditions. This is necessary because the operating parameters of silicon devices are dependent on temperature, and this dependence cannot be extrapolated from foundry-provided models (often only rated down to -55 °C \cite{10.1117/12.2219734}) due to the onset of cryogenic effects which contribute significantly to very low-temperature behavior. The device parameters captured in the process design kit (PDK) used in electronic design automation (EDA) tools need to accurately reflect realistic operation at low temperature. Significant efforts to characterize and model the behavior of low-temperature transistors has been performed \cite{BeckersJEDS2018,BeckersSSE2019,9265034,9503117}.

\jonE{The full range of device variability that can be expected due to factors such as random dopant fluctuations, variability in critical device dimensions, and variations in material purity must also be captured. The effects of this variation on device behavior changes and often increases from room temperature to cryogenic temperatures~\cite{9072133}; therefore, multi-terminal current-voltage (I-V) characterization of a large number (thousands) of individual transistors at cryogenic temperature is required to extend the simulation of such transistors into the cryogenic environment.} 

\jonE{Capturing a comprehensive set of cryogenic test data presents logistical challenges. The overhead of wire-bonding individual transistor device terminals combined with practical constraints of die size, cryostat wiring, and thermal cycle times make this approach impractical. At room temperature, these measurements are typically performed at the wafer scale, using an automated wafer prober to access dedicated pads for each transistor across large arrays of devices. There are advantages to this approach. Wafer-level probing of individual devices allows for simpler and more standardized de-embedding approaches to be followed}. It has been shown that this method can be extended to cryogenic environments by utilizing highly specialized cryogenic probe systems~\cite{neyens2023probing}. This method allows wafer-scale variability to be measured in an analogous way to the traditional foundry PDK development process; however, it lacks flexibility and accessibility due to the heavy dependence on expensive, specialist equipment and inefficient use of wafer area.


An alternative is to use an integrated multiplexer to switch multiple devices to a single set of pads on a die for analog force and sense connections. \jonE{This greatly increases the number of devices that can be characterized in a set time frame and die area compared to bonding individual transistors while greatly reducing the cost-of-entry barrier compared to full wafer probing. Multi-project wafers (MPWs) can be used for fairly cost-efficient fabrication of a small number of test dies without paying for a full wafer. Still, to incorporate cryogenic variability on a global scale (inter-die, inter-wafer, etc.) into a PDK in addition to local mismatch, multiple dies must be measured, so a bonded-die-based measurement system will benefit from the ability to easily swap bonded samples while minimizing disturbance to the rest of the measurement system.}

Here we report measurement results from such a system constructed for the purpose of testing cryogenic silicon components at the lowest level (i.e. individual transistors) in support of cryogenic PDK development. In this system, a combination of on-die multiplexers and a modular daughter board/socket configuration for bonded samples grants a large degree of flexibility. Thus, the same core set-up can be used across standard research-focused cryogenic systems such as a dilution fridge or a dipping probe in a helium dewar, and bonded samples can be rapidly swapped out limited only by the time needed for thermal cycling of the system.

\section{Requirements}
Here, we briefly outline the requirements for the test system.

\textit{Temperature: } Positioning of the various elements of control and readout circuity in a quantum computing system is based on a compromise between the benefits of compact integration with the quantum device layer and cooling power requirements \cite{xue2021cmos}. It would be more convenient to operate quantum devices and all classical control/readout electronics at the same stage; progress on the operation of spin qubits at elevated temperatures \cite{yang2020operation,camenzind2022hole,ono2019high} may permit a single common cryogenic stage operating in the few-Kelvin regime. Low-power multiplexer circuits and their associated digital control circuitry are well suited for monolithic integration with quantum devices to reduce the off-die wiring requirements without raising the on-die temperature to the point of significant detrimental effects on spin qubits \cite{yang2020operation}. Some control circuitry such as simple signal generators may not exceed the cooling power limitations of the mixing chamber plate of a dilution fridge but may be better suited to heterogeneous integration in which these electronics are on a separate die from the quantum devices and are largely thermally isolated to avoid degrading qubit fidelities \cite{pauka2021cryogenic}. More power-hungry signal chain components such as amplifiers, mixers, and analog-to-digital converters (ADCs) typically must be operated at higher-temperature stages (e.g. the second stage of the pulse tube cooler of a dry dilution fridge sometimes referred to as the PT2 stage, typically $T\approx3.5$~K) where cooling power is much greater but latency and wiring are still minimized, and cryogenic performance benefits can be realized. In any case, one must perform characterization measurements in a system with very similar temperature and thermalization profile to that of the intended end application to ensure applicability and accuracy of any derived models. Local on-die thermometry is also required to fully understand the effects of self-heating and heating among devices on the same die.

\textit{Bias/power lines}: Standard integrated circuit design best practices such as electro-static discharge (ESD) protection and digital logic reset control are required to produce robust and reliable parts. Additionally, devices with different maximum operating levels are typically  used (e.g. thin- and thick-oxide gates for core and IO implementation, respectively). As such, multiple supply voltages need to be delivered to the die. Additionally, for silicon-on-insulator (SOI) technologies, backgate bias voltages for some of the circuitry may be externally driveable to allow for tuning the performance~\cite{han2022back}. As the device operating conditions can vary drastically between room temperature and cryogenic temperatures, the instruments supplying these voltages must have current-limiting/-monitoring and appropriate power-up sequencing capabilities to ensure proper operation and optimize for low power. 

\textit{Digital control}: Logic signals are required to initialize the state of multiplexers and other configurable components of the logic circuit. The maximum clock rate of these signals may be limited by the use of low-dissipation cryogenic wiring, which is often in the form of lossy resistive twisted pairs with no individual screening. However, operation at slower digital clock rates (e.g kHz range) to accommodate such wiring is typically acceptable, as the digital instructions to enable multiplexer connections to a device can still be delivered in a timescale that is orders of magnitude shorter than the measurement time for that device.

\textit{Packaging}: A demountable sample package is required for quick and easy swapping of pre-bonded samples. This must also be compatible with ESD requirements.

\textit{Instruments}: Two low-noise, high-resolution source-measure units (SMUs) are required to characterize drain current and gate leakage current while a third SMU maintains a common potential for the source terminal of the device independent of voltage drops across wiring/multiplexer terminals. Note that in this paper, DUT refers to the entire die, whereas individual transistor devices being measured on the die are simply referred to as devices.

\section{Measurement System}

\begin{figure}
    \centering
    \includegraphics[width=1\linewidth]{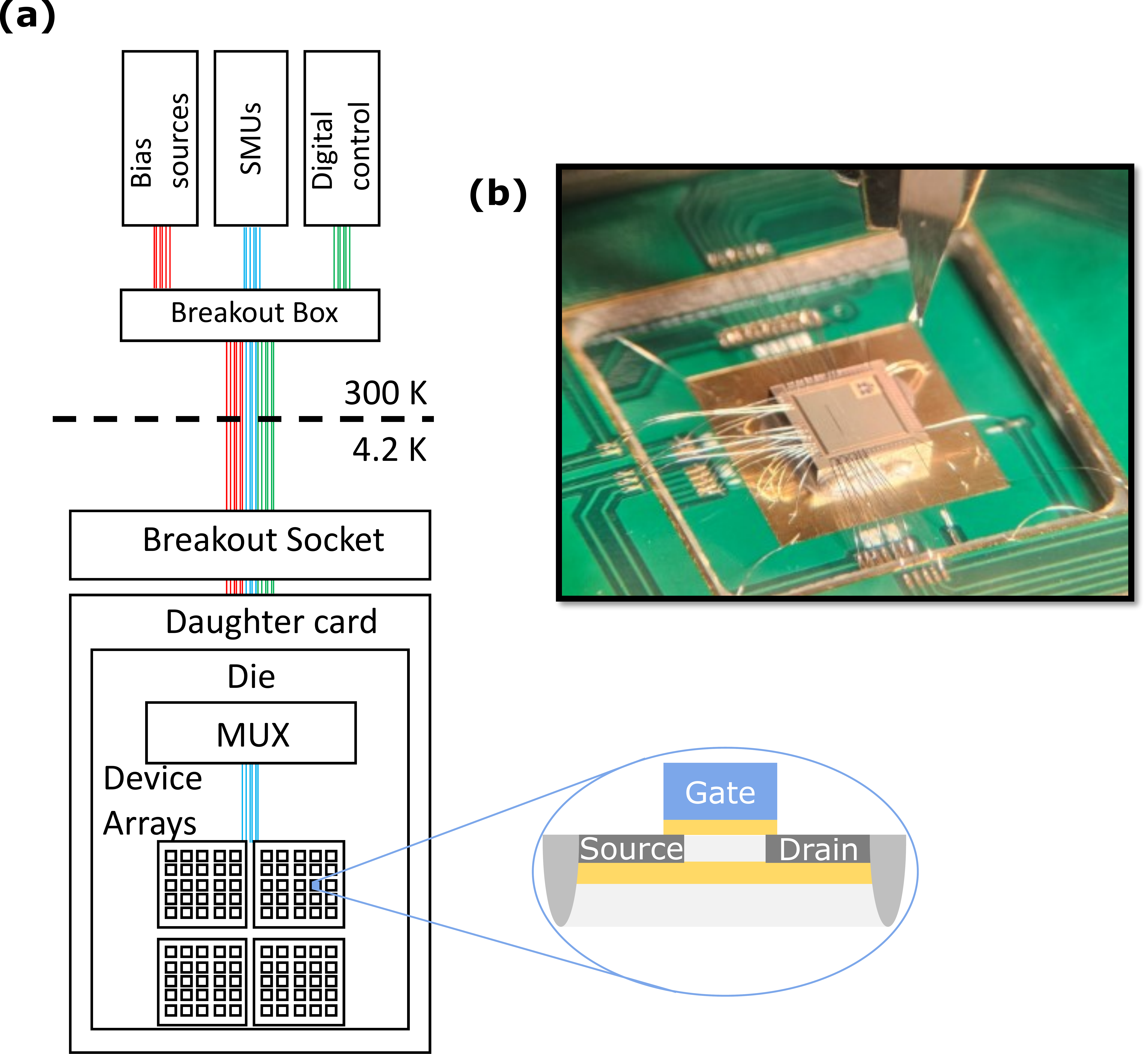}
    \caption{\jonE{(a) Block diagram of the key components in the measurement system. A die containing multiple device arrays, each consisting of 1024 transistors accessed via a multiplexer is wire-bonded to a demount-able PCB "daughter card." A corresponding multi-pin socket mounted to a liquid helium dipping probe provides an interface to the digital circuitry necessary to configure the state of the multiplexer as well as the analog circuity used to conduct the measurements on  the selected device. (b) Photograph of the bonded die. By recessing the die within a multi-level PCB and utilizing the surface of both levels, the density of bond pads can be increased, minimizing the length and approach angle of bond wires needed.} }
    \label{fig:figsystem}
\end{figure}
\subsection{Overview} 

We have created a measurement system for the sensitive DC analog tests of multiplexed cryogenic components. Figure \ref{fig:figsystem}(a) shows a block diagram of the measurement system. At room temperature three distinct types of instrumentation are used: DC bias sources, digital pattern generators, and SMUs. These are coupled to the DUT using a cryogenic wiring loom. We use a dipping probe for immersion of test samples into a liquid helium bath (T = 4.2~K). This provides a rapid turnaround for sample testing at a similar temperature to that found on the second stage pulse tube cooler of a dilution fridge (T = 3.5~K). This ensures excellent thermalization but does not easily allow for measurements at temperatures other than that of liquid helium, making development of an isothermal model the most straightforward option. Isothermal modelling is discussed in more detail in section IV. D. The sample is wire-bonded to a daughter board which mates with a cryogenic socket. The DUT on the daughter board could be any integrated circuit with separate terminals for bias, digital and analog lines, extending the use of this system beyond just individual transistor measurements.

\subsection{Integrated Circuits (DUTs) and Sample Packaging}
Two integrated circuits designed in a 22-nm UTB SOI process are measured, each with a similar implementation of integrated multiplexers to access thousands of devices for measurement along with on-die digital control logic and ESD protection. An example sample is shown in Figure \ref{fig:figsystem}(b) adhered and bonded onto a custom daughter board. A multi-layer sample mount FR4 PCB is used to provide optimized placement schemes for the $\sim$50 wire-bonds required to a DUT. A manual wire bonder is used with Al:Si (1\%) wire of diameter either 25 $\mu$m or 17.5 $\mu$m depending on the DUT. A multi-pin connector socket on the cryogenic system interfaces with contact pads on the daughter board.

Each die contains four 1-to-1024 multiplexers, each connecting to a "farm" containing a different family of transistor devices which can be individually accessed for a total of eight transistor families able to be modeled. Each of these farms is in the form of 32 by 32 arrayed individual transistor devices of varying width, length, and number of fingers (NF) to allow for model extraction. In the second integrated circuit, a few positions in each farm are occupied by short, open, and 50 $\Omega$ load structures to allow for multiplexer characterization and some de-embedding. A trade-off between having more copies of same-size devices for variability analysis and having a greater variety of different dimensions present within the farm for modelling of real device geometry-scaling effects must be considered when determining the farm contents. 

The core multiplexer unit cell structure contains an analog signal pass gate, or switch, formed by an N-channel FET (NFET) and P-channel FET (PFET) in parallel with digitally driven gate voltages to define the on and off states as shown in Figure \ref{fig:circuit_diagrams}(a). Since the increased threshold voltages of these devices at cryogenic temperatures result in substantially increased on-resistance when the pass voltage is near the midpoint of the high and low supply voltages, $V_{DD}$ and $V_{SS}$ (in some cases both the NFET and PFET may be in the subthreshold regime even with the pass gate in the "on" state), the backgate voltages (not shown) of the multiplexer pass gate FETs are externally driven to enable shifting the device thresholds nearer to their room-temperature values~\cite{han2022back}. 

All pass gates in the four farm multiplexers on a die share the same NFET backgate bias and the same PFET backgate bias voltage since individual control is not practical. Each transistor device has 2 pass gates connected to each of its source, gate, and drain terminals with a star connection as close as possible to the device to allow for local voltage sensing in a Kelvin measurement setup for each terminal, compensating for DC voltage drop across the multiplexer and the rest of the inline system components. The careful routing of this sense line connection is vital for devices that support high-current operation. An additional 2 pass gates are connected to gate contacts on the opposite side of the channel to allow for local gate resistance thermometry (GRT)\cite{noah2023cmos}, allowing for characterization of self-heating of the devices. Thus, a total of eight bond pads can be used to access a single farm, and 32 total bond pads allow complete access to the four farms on a die.

The die contains a file of 32-bit registers such that a register for one-hot row selection and another register for one-hot column selection are used to turn on the pass gate to the device at the selected row/column pair. An on-die test access port (TAP) compliant with IEEE 1149.1 \cite{jtag} allows for writing and reading the contents of the registers via a standard digital interface supporting up to MHz-range clock rates.


\subsection{Wiring and Continuity Tests} 

Enamel-coated cryogenic wiring looms (twisted pairs) were used\cite{cunningham1995woven} within the dipping probe. Two looms of 12 pairs each provide enough wiring for several sets of force/sense lines along with biasing voltages and digital control lines. This form of wiring controls the thermal load into the dewar and gives reasonable protection from vibration-driven triboelectric noise which can affect low-level current measurements~\cite{kalra2016vibration}. Using matched conductor material (and hence Seebeck coefficient) on pairs of analog lines nominally cancels any thermoelectric voltages across the temperature gradient. The lack of shielding is not a problem during measurements (when the digital bus is quiet) but this does preclude very high frequency operation for digital lines.

After preparing the sample and again after cooling, a test procedure can be performed to test the connectivity of the wiring using the expected properties of the ESD protection circuits on the various lines. The characteristics of the ESD protection diodes change with temperature such that high voltages may be reached following typical constant-current continuity test routines at cryogenic temperatures - care should be taken to minimize the applied bias magnitude and duration to avoid damage to internal circuitry while maintaining the ability to discern the diode response in these conditions. These measurements can be used to corroborate the successful connection of each wire to an active pin and detect rare failures of the bonding that can occur during thermal cycling.

\subsection{System Bias and Digital Instrumentation}

Bias signals are sourced from a room-temperature multi-channel DAC with per-channel current monitoring. The current draw on each line is a key diagnostic that can be used to inform a software limit on power-up. Power-up sequencing and slew rate limiting rules are followed to prevent high transient currents and ensure the DUT is repeatably initialized into the desired state. Digital logic signals at kHz-range frequencies are sufficient to address and write the multiplexer control registers without a major impact on overall measurement time. A room-temperature opto-isolated digital pattern generator and logic analyzer along with a logic-level shifter are used to transmit and receive digital communications with the DUT.

\subsection{Analog Instrumentation}
\begin{figure}
    \centering
    \includegraphics[width=1\linewidth]{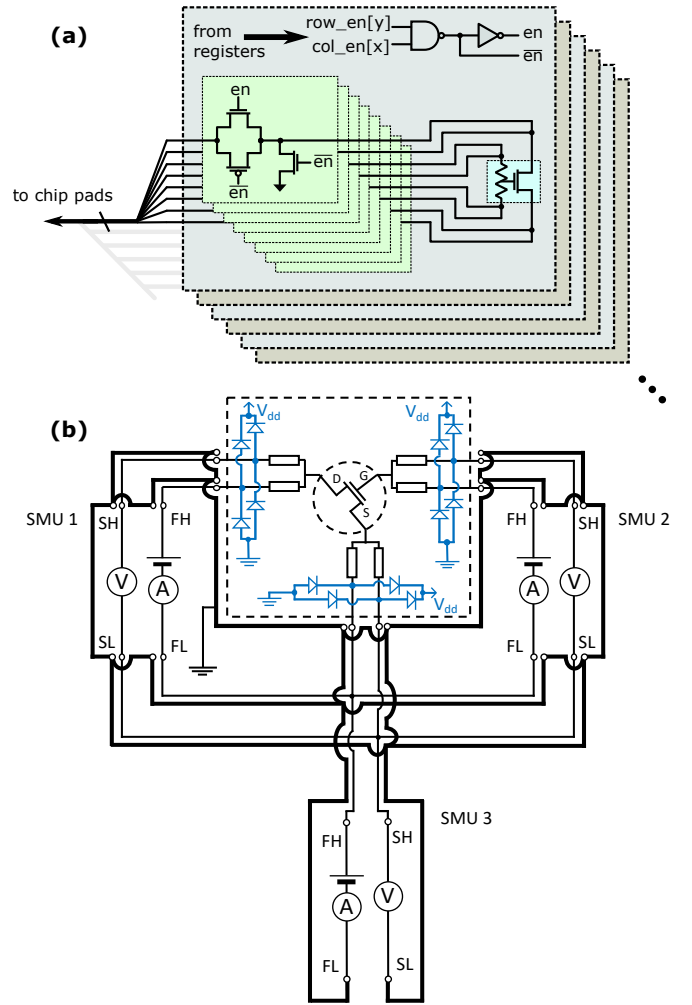}
    \caption{(a) High-level circuit diagram of an on-die multiplexer structure allowing analogue connections to be established with Kelvin sense to each terminal of a selected transistor. Gate contacts are established at opposite sides of the channel to allow gate resistance thermometry to be performed. A pair of 32-bit registers (one for one-hot row selection and another for one-hot column selection) are used to turn on the pass gate to the device at the selected row/column pair. Each signal pass gate is formed by an NFET and PFET in parallel. For a farm containing NFET devices for measurement as shown, a pull-down FET is activated for devices that are not selected; for a farm containing PFET devices for measurement, a pull-up FET is used instead. (b) Diagram of analog instrument connections used to conduct a 6-terminal measurement with Kelvin sense (FH, SH, FL, and SL corresponding to force-high, sense-high, force-low, and sense-low channels of SMUs, respectively). The ESD protection diodes in blue demonstrate the value of using SMU3 as a reference to the system ground. SMU3 in this case can bias the source at an arbitrary reference potential with respect to system ground, enabling full control and optimization of measurements to avoid excess forward-bias currents flowing in any ESD diodes while maintaining high-impedance sense lines.}
    \label{fig:circuit_diagrams}
\end{figure}

Analog transistor DC characteristics can be probed with multiple source-measure units (SMUs) as shown in Figure \ref{fig:circuit_diagrams}(b). Typically, one unit controls drain-source bias and the other controls gate-source bias. These are bound to a common reference point whose potential can be arbitrarily set with respect to chassis ground by the third SMU. When using a Kelvin sensing configuration, each device terminal must have independent lines for current injection ("force") and voltage monitoring ("sense").  These can be implemented as parallel switching routes to a multiplexed device terminal as previously described. Sensing voltages on the device side of the multiplexers allows for accurate real-time compensation for voltage drops due to the impedance of the measurement leads and multiplexer switches. This system also gives control over how the analog circuitry as a whole is referenced to the low-side supply $V_{SS}$. This allows the full range of accessible potentials on each terminal, defined by ESD protection structures, to be utilized.

\jon{The drain-source current in a gate sweep will span a range of sub-nA currents in the subthreshold regime to more than 0.1~mA in inversion. This is often best handled with a fixed current measurement range to avoid discontinuities at range changes, though this may sacrifice precision in the subthreshold regime.} The accuracy of SMU voltage output can be checked against an instrument with a calibration traceable to the SI Volt via the Josephson effect~\cite{NPLVolts}. Using an HP-3458A digital voltmeter (DVM) as a transfer standard, with accuracy better than $1$~part per million (ppm), we have checked that the deviation in the voltage sources is $<200~\mu$V/V in the range $0-2$~V, within the accuracy quoted by the manufacturer of 500~$\mu$V/V at 1~V~\cite{K2450Datasheet}. The current measurement accuracy of each SMU can be established by setting its output to 0~V and measuring a reference current generated by applying a known potential difference across a standard resistor whose value (known better than $0.5$~ppm) is traceable to the quantum Hall effect~\cite{NPLResistance}. The current readout gives an uncertainty within $10$~ppm of nominal, within the manufacturer's specification~\cite{K2450Datasheet}. This performance is adequate to extract parameters like on current $I_{d ON}$ and the threshold voltage $V_{th}$ at the $<1$\% level as seen below without further corrections.

\subsection{Leakage Subtraction}

\begin{figure}
    \centering
    \includegraphics[width=1\linewidth]{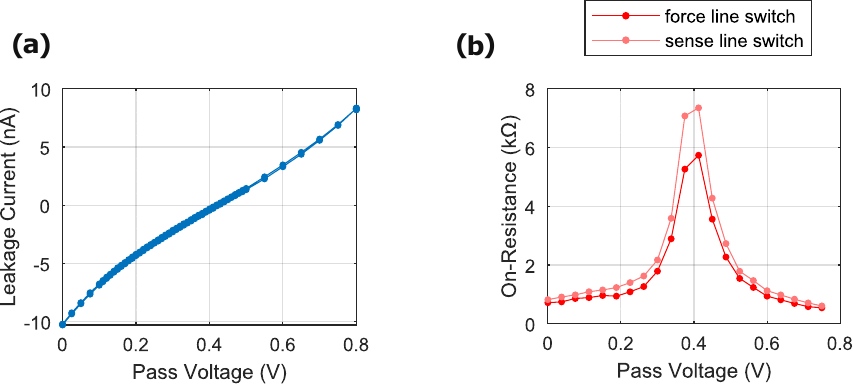}
    \caption{Example 0.8-V farm multiplexer switch characteristics measured at 4~K across a range of pass voltages (relative to the ground voltage of the switch circuitry) with appropriate switch FET backgate biases applied to enable cryogenic operation. (a) Leakage current measurement conducted on an open structure, including leakage current effects from both the force line switch and the connected sense line switch. (b) Force line and sense line switch on-resistances independently measured at 1 \textmu A pass current.}
    \label{fig:muxchar}
\end{figure}

In an accurate electrical measurement, even sub-nA leakage between terminals can result in measurement errors. This can arise in PCBs, room-temperature and cryogenic cabling, and on the die itself. The leakage in the wiring can be tested by removing the sample assembly. The leakage of cryogenic loom varies between manufacturers and is also temperature dependent, typically reducing markedly at low temperature. A wire-to-wire leakage of $> 100~{\rm G}\Omega$ is desirable.

Leakage on the die arises mainly from ESD protection structures, off-state drain-source leakage through the disabled pass FETs of other multiplexer channels, off-state leakage through the disabled pull-down/-up FETs, and gate leakage from the various FETs. These leakage effects can be measured (and then subtracted from actual device measurements) by measuring the open structures in the farms. As leakage current is often voltage-dependent, one must take care to apply the best approximation of voltages the different FETs will experience during device measurement when performing these leakage measurements for subtraction. The sense lines and their switches are designed for minimal leakage such that leakage due to the disabled pull-down/-up FETs are able to compensated. An example open structure leakage current measurement (including leakage from both force and sense switches) along with separate force and sense switch on-resistances are plotted across switch pass voltage in Figure~\ref{fig:muxchar}.


\section{Results}

\begin{figure*}
    \centering
    \includegraphics[width=1\linewidth]{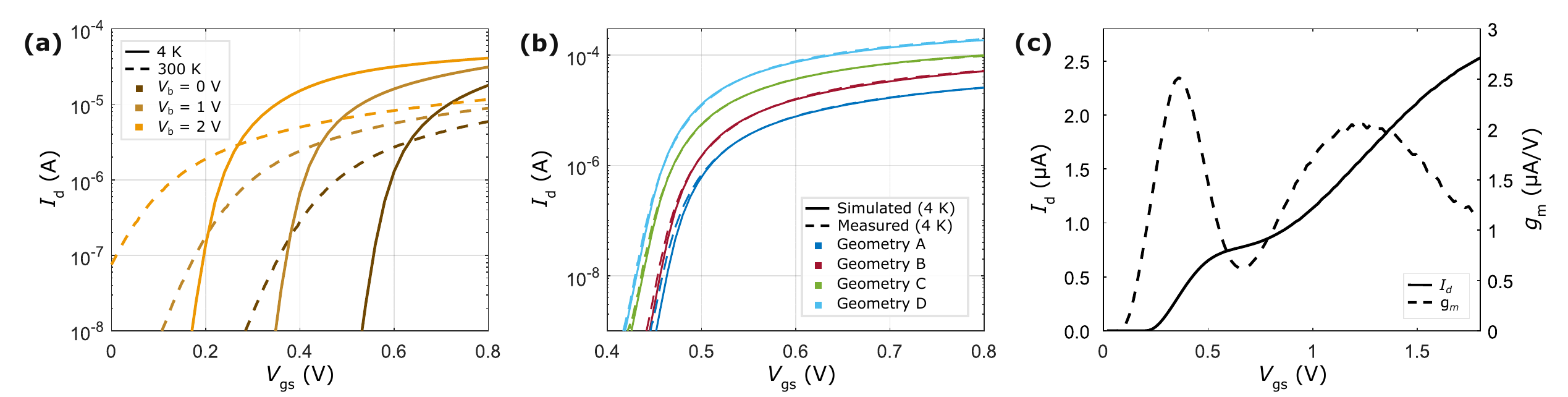}
    \caption{\jonE{(a) Drain current of a single NFET as a function of gate-source bias ($I_{d}$-$V_{gs}$ curves). Measurements were taken at $V_{ds}$ of 0.05 V with backgate voltages ($V_{bs}$) of 0 V, 1 V and 2 V (increasing backgate voltage corresponding to lighter shade). Measurements taken at 300 K and 4 K can be seen with solid and dashed lines, respectively.} (b) $I_{d}$-$V_{gs}$ curves at $V_{ds}=0.05$ V and backgate voltage  of 0 V at 4 K for four geometries of NFETs represented by different colors. Dashed lines represent measurement data, while solid lines represent BSIM-IMG v102.9.6 model simulations using model parameters extracted from the 4 K measurement data demonstrating a relatively good fit as a proof-of-concept for cryogenic PDK model building. (c) $I_{d}$-$V_{gs}$ curve (solid line) at $V_{ds}=0.05$ V and backgate voltage of 2 V at 4 K for an NFET demonstrating the intersubband scattering effect. The corresponding transconductance ($g_{m}$) curve is smoothed and plotted as a dashed line to further display the characteristic behavior of this effect.}
    \label{fig:figdata}
\end{figure*}

\subsection{Representative I-V Curves}

\jonE{Figure \ref{fig:figdata}(a) shows example drain current vs gate-source differential voltage ($I_{d}$-$V_{gs}$) curves from a single device at 0.05 V drain-source bias ($V_{ds}$) at a range of backgate voltages at temperatures of both 4 K and 300 K. Key parameters such as the threshold voltage, subthreshold slope, and mobility can be extracted as a function of backgate voltage, drain-source bias, and temperature. } Data taken under various bias conditions can be used to extract DC parameters for an industry-standard compact model such as BSIM-IMG (a multi-gate model applicable because the transistors are in an SOI technology with a backgate)~\cite{bsimimg}. Fitting measured data across a range of device geometries yields a model card (i.e. a set of model parameters used for simulating a certain device type) that can then be used for arbitrary DC simulations of any device size and bias conditions. A demonstration of simulations using a BSIM-IMG v102.9.6 model card extracted from measured 4~K data and closely matching the corresponding measurements is shown in Figure~\ref{fig:figdata}(b) for four geometries at $V_{ds}=0.05$ V.

\subsection{Parameter Selection and Sweep Rates}

\dd{For DC I-V characterization at cryogenic temperatures, the optimal step size within a voltage sweep considers the trade-off between measurement time and the accuracy in an extracted model's I-V curve shape. As a rough estimate, measurements at approximately $40-100$ gate-source biases are typically required for a $V_{gs}$ sweep extending from the subthreshold region into the strong inversion regime to provide sufficient information for model building; fewer points are generally required for $V_{ds}$ sweeps. Careful determination of threshold voltage and transition between different regions of operation is crucial for model parameter extraction and validation. In these regions in which the real device effects of channel length modulation, drain-induced threshold shifts, etc. \cite{ch3} can be observed, smaller steps may be justified. In the deep-subthreshold region, FET devices also show gate-induced drain- and source-lowering effects \cite{ch4}. Due to the very low current values, modelling such regions of operation requires denser sweep points. However, the kind of circuit to be designed (e.g. digital, high-power vs. low-power analog, RF) must be considered. For some applications, very low current regions of operation may not require a very accurate model and therefore a coarse sweep for this region of operation will suffice. For digital circuits the threshold voltage and the current in saturation are critical, so step size may be decreased in those regions of operation to ensure an accurate model.}


\dd{Typically, curve shapes in moderate and strong inversion regions show predictable behaviors that can be fit well from relatively coarse sweep points; however, a few exceptions exist for cryogenic behavior. For example, certain devices exhibit an intersubband scattering effect in the inversion regime requiring a smaller step size in gate voltage sweep to determine a good fit for the observed behavior as shown in Figure~\ref{fig:figdata}(c). This behavior results from a transition from conduction within a single energy subband in weak inversion to two-subband conduction as higher inversion carrier density in the lowest subband leads to some carriers beginning to occupy the second lowest subband \cite{9204657}. This results in a sharp decrease in mobility as scattering between the two energy levels of electrons can then occur, producing a hump in drain current $I_{d}$ and a double hump in transconductance $g_{m}$. Industry-standard compact models such as BSIM-IMG do not currently include the intersubband scattering effect, and as such, cryogenic PDK developers must create custom compact model equations for this behavior \cite{aouad:tel-03852447}.}


The time limitation imposed by the cryogenic measurement setup is a critical consideration to ensure the devices under test remain under constant ambient temperature. "Wet" cryogenic dipping systems can support experiments of at least a few weeks, while dry systems can run continuously. \dd{For DC characterization the final settling values of current and voltage need to be measured. A delay between bias application and measurement is necessary to allow sufficient settling time. Settling time estimations can be made by analysing the effect of path parasitic elements in the system and/or experimentally determined. After the settling time, the measurement should be integrated over an integer multiple of power line cycles (PLC) to integrate out any effects of the power line frequency variations coupled to the measurement instruments. For 50 Hz power line frequency as in these measurements, the minimum integration time of 1 PLC is thus 20 ms.} It is also beneficial to conduct each parameter sweep in both directions; in addition to highlighting subtleties in the measurement system/routine such as insufficient settling time, this can be a vital tool in attributing unexpected features to device physics, self-heating, or other hysteresis. One could also collect the measurement points in a random order to evaluate any device memory effects.


\subsection{Bulk Measurements}

\begin{figure*}
    \centering
    \includegraphics[width=0.9\linewidth]{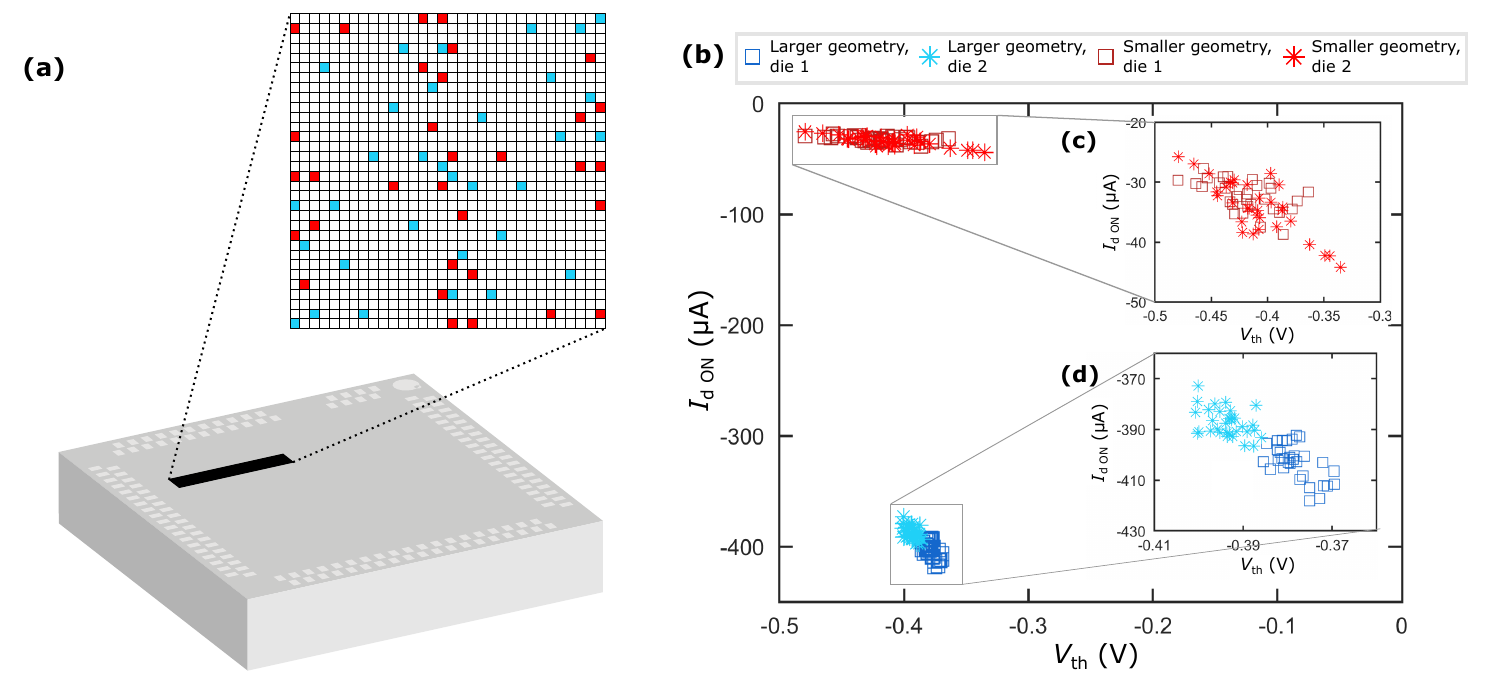}
    \caption{\jonE{ (a) Depiction of a die containing device farms and integrated multiplexer, highlighting the location of one of the farms. Each farm consists of a 2D array of transistors of various geometries in a pseudo-random common-centroid configuration.  The relative position of PFET devices with 2 nominal geometries are represented by red and blue squares (smaller and larger dimensions, respectively). (b) Parameter spread at $V_{bs}=-0.8$ V of threshold voltage ($V_{th}$) and on current $I_{d ON}$ ($V_{gs}=-0.8$ V and $V_{ds}=-0.8$ V) of devices with the corresponding color in (a). The geometry represented by the blue points has a greater width-to-length ratio, explaining the greater magnitude of $I_{d ON}$. Measurements were obtained across two nominally identical dies (indicated by square and asterisk point markers. Insets (c) and (d) each highlight the parameter spread of devices within a single nominal geometry.}}
    \label{fig:device_comparison}
\end{figure*}

Access to a multiplexed array of devices enables a detailed systematic study of device design choices and variability of device behavior. \jonE{ Figure \ref{fig:device_comparison}(a) highlights the location of one of the farms on a die. Each farm contains multiple copies of a number of distinct geometries. Devices are distributed in a pseudo-random configuration with each set of same-geometry devices arranged to have its centroid at the center of the farm to reduce systematic perturbations in device parameters due to relative layout effects. Figure \ref{fig:device_comparison}(b) shows two key characteristics ($I_{d ON}$ and $V_{th}$) across 2 geometries and across 2 dies, indicating the scale of variation observed across nominally identical devices.} 

\gn{To capture inter-die variation effects, repeating measurements across a random sample of a number of dies can begin to give a picture of the distribution of device behavior that may be present. This variation may represent intra-wafer, intra-lot, or even global variation for a process, depending on how the die samples are fabricated/selected. For devices with smaller dimensions, the global variation is relatively small compared to the local mismatch variability which can be observed among nominally identical devices on the same die. This is demonstrated in Figure~\ref{fig:device_comparison}(c), in which a small device geometry demonstrates high variability across each die with both dies' distributions largely overlapping. For larger devices, global variation effects can play a bigger role than local mismatch effects. A larger device geometry exemplifies this in Figure~\ref{fig:device_comparison}(d), where the difference in average characteristics across two dies is of similar magnitude to the maximum intra-die variation on each die.} The parameter variation seen across Figure \ref{fig:device_comparison} exceeds the distribution from measurement uncertainty, and the visible variation at this scale is dominated by device variation.

Statistical data for each distribution in Figure~\ref{fig:device_comparison} can be found along with comparisons to their simulated room-temperature counterparts in Table I. While the level of variability observed in the smaller geometry is much greater than that of the large geometry, it is actually the larger geometry whose relative increase in variability compared to room-temperature values is more significant. In fact, for one of the dies, the observed cryogenic variability of the smaller geometry (which represents very nearly the minimum drawn dimensions available for a PFET in the studied technology node) is actually lower than its predicted room-temperature variability. This underscores the need for large sample sizes of devices and multiple dies to be characterized to draw accurate conclusions about population variability. The relative (percent) increases in parameter sample standard deviations from room temperature to 4~K are overall comparable to published cryogenic variability increases in other technology nodes \cite{9265034, 9072133}. This increase in variability is critical to characterize as part of cryogenic PDK development to allow circuit designers to make informed decisions about performance and variability trade-offs and to determine what kind of trim capabilities may be needed to meet desired performance specifications for a given circuit.


\begin{table}
\caption{\label{} Sample statistics of extracted parameters for selected PFET geometries at 4~K with room-temperature comparisons
}

{\renewcommand{\arraystretch}{1.4}
\begin{tabular}{>{\hspace{-2.5pt}}l<{\hspace{-2.5pt}}|>{\hspace{-2.5pt}}c<{\hspace{-2.5pt}}>{\hspace{-2.5pt}}c<{\hspace{-2.5pt}}>{\hspace{-2.5pt}}c<{\hspace{-2.5pt}}>{\hspace{-2.5pt}}c<{\hspace{-4pt}}}
\hline
\hline
 & Larger & Larger & Smaller & Smaller \\
 & geometry, & geometry, & geometry, & geometry,\\
Parameter  & die 1 & die 2 & die 1 & die 2\\
\hline
Sample size & 32 & 32 & 32 & 32\\
\hline
4K $\overline{V_{th}}$ (mV) & -378.2 & -393.7 & -420.2 & -410.0\\
4K $\sigma(V_{th})$ (mV) & 4.3 & 4.2 & 27.1 & 33.1\\
$\underset{\textrm{RT}\rightarrow\textrm{4K}}{\Delta}[\overline{V_{th}}]$ (mV)* & -135.6 & -151.0 & -141.8 & -131.6\\
$\underset{\textrm{RT}\rightarrow\textrm{4K}}{\Delta}[\sigma(V_{th})]$ (\%)* & 66.5\% & 63.8\% & 2.7\% & 25.3\%\\
\hline
4K $\overline{I_{d ON}}$ (\textmu A) & -403.1 & -387.5 & -32.2 & -34.1\\
4K $\sigma (I_{d ON}/\overline{I_{d ON}})$ (\%) & 1.7\% & 1.4\% & 7.9\% & 13.4\%\\
$\underset{\textrm{RT}\rightarrow\textrm{4K}}{\Delta}[\overline{I_{d ON}}]$ (\%)* & -1.0\% & -4.9\% & -18.3\% & -13.5\%\\
$\underset{\textrm{RT}\rightarrow\textrm{4K}}{\Delta}[\sigma (I_{d ON}/\overline{I_{d ON}})]$ (\%)* & 74.8\% & 43.1\% & -17.6\% & 39.8\%\\
\hline
\end{tabular}
}\\\\
*Delta parameters represent 4-K measurements' difference from room-temperature Monte Carlo simulations of local mismatch variation (sample size of 100) using foundry-provided model cards and taking into account local layout effects and parasitic extraction for the devices in the farm.

\label{tab:stats}
\end{table}


To support high-quality model-building, $V_{gs}$ sweeps must be conducted at several $V_{ds}$ bias voltages, and $V_{ds}$ sweeps are often conducted at a few $V_{gs}$ bias voltages. Then this entire set of $V_{gs}$ and $V_{ds}$ sweeps are carried out at multiple backgate bias voltages. Thus, we can estimate total measurement time for a farm as a multiplication of several key chosen parameters as given in Table II. The total number of measurement points per device is of order $10^3$. For a farm of 1024 devices, this gives a total of approximately $10^6$ data points per farm. Using the values in Table II gives a minimum raw measurement time of at least 7.5 hours not including any overheads. The total time taken for collection of a full measurement set is greatly increased by a characterized settling time requirement of $\geq 20$~ms. Accounting for this set-measure delay along with instrument communication time related to multiplexer selection, actual measurement time is closer to 20 hours. Optimising the multiplexer selection time (e.g. via digital pattern pre-loading) would make only a small difference, reducing the total time to a minimum of $\sim19$ hours. A better route to reducing total measurement time is to reduce the required set-measure delay, which requires further optimisation of wiring both on-chip and in the probe. Additionally, if a sufficient number of SMUs are available, a device in each of the four farms on a die may be measured simultaneously in parallel, drastically reducing the effective measurement time per device. However, one must consider the effects of cross-device heating on the same die may be significant. Furthermore, multiple dies may be connected in such a way that they share power supply lines and digital control signals (but separate analog connections to their farms) to allow parallel measurements across several dies at once, providing further potential orders of efficiency improvement.
%



\begin{table}
\caption{\label{} \jonE{Representative time budget for a full measurement set across a 1024 device transistor farm.
}}
{\renewcommand{\arraystretch}{1.2}
\begin{tabular}{lcr}
\hline
\hline
Parameter & Quantity & Unit\\
\hline
Test devices & 1024 & per farm\\
Sweep length &  55 & points\\
Sweeps per backgate bias voltage & 8 & sweeps\\
Backgate bias voltages & 3 & points\\
Integration time & 20 & ms\\
\textit{MUX selection time (per device)} & 5 & s\\
\textit{set-measure delay (per point)} & 30 & ms\\
\hline
Total raw measurement time & 7.5 & h\\
\hline
\textit{Total measurement time} & 20.2 & h \\
\hline
\end{tabular}
}
\label{tab:timeBudget}
\end{table}

Recently, a method has been shown which enables rapid measurements of multiplexed silicon CMOS transistors operating in the regime dominated by single-electron transport\cite{thomas2023rapid}. This is proof of principle for a time-domain multiple access (TDMA) scheme for quantum devices. A single RF reflectometry readout line measures the changes in device impedance associated with the formation of Coulomb blockade state, i.e. where electron transport is inhibited by the energy cost of adding a single electron to an isolated reservoir~\cite{vigneau2023probing}. That work described a similar farm being measured within a matter of minutes rather than hours, though the goals of the two measurements are not comparable. This DC transistor characterization work takes significantly longer due to the reduced bandwidth of DC measurements, but such measurements are required for accurate extraction of DC device characteristics for PDK development - reflectometry would not provide sufficient information.

\subsection{Isothermal Modeling}

An attractive approach to deep-cryogenic transistor model-building for quantum computing applications is to perform all measurements near 4~K (as opposed to lower temperatures) where material thermal conductivities are higher, promoting better thermalization, and relatively high cooling power helps ensure stable operation to build reliable isothermal 4~K models. Isothermal models theoretically do not require measurements at higher ambient temperatures to characterize detailed temperature dependencies since the local self-heating effects are included in the core isothermal model itself~\cite{workman1998physical}. Transistor behavior below 4~K shows little change compared to 4~K behavior due to a combination of physical parameter saturation and local self-heating during operation\cite{9265034}. Therefore, such models would then also be able to quite accurately predict transistor behavior at any ambient temperatures below 4~K.

\gn{If the measurement data for isothermal model building is collected in a liquid helium immersion environment, then the self-heating effects will be different from those in a vacuum (e.g. dilution refrigerator). This means if an isothermal model built from liquid-immersion measurement data is used to simulate transistors operating at high current in a dilution refrigerator, then some aspects of the behavior will not be accurately simulated. In a vacuum, transistors under the same bias conditions will generally be at a higher temperature due to comparatively poor thermalization. Thus, for applications where a significant level of current will be passed through transistors, it is especially important for the model-building measurement environment's thermalization profile to match the intended application environment.}

\dd{A limitation of an isothermal model is its limited capability to capture a device's local heating effect when a large transient signal is applied, e.g. in the output drive stage of amplifiers or the rail-to-rail swing of buffer stages. Even if the ambient temperature is constant, the local heating of a FET transitioning from subthreshold to strong inversion can be different compared to the reverse transition. The effect of the surrounding structures, device layout, and operating frequency may add further uncertainty to the device's behavior. Even with this limitation, the isothermal modeling approach gives a meaningful physical understanding of the device's operation. Under very low-current device operation and sufficient cooling capacity, one can assume the device core temperature to be the same as ambient. This can be utilized to capture certain key aspects of device behavior in the model card. Hence, isothermal modeling can be completed as a crucial step toward an improved modeling approach that includes device dynamic heating effects.}

\subsection{Other Measurement Types and PDK Considerations}

\gn{We have discussed a system which can fully meet the needs of DC I-V characterization for PDK development; however, additional measurements are required to enable accurate RF simulations in a PDK. These high-frequency S-parameter measurements are somewhat limited by the presence of the multiplexer structure compared to direct measurement of transistor devices. Some de-embedding is possible in the farms as previously described, but only reasonably in sub-GHz frequency ranges and even then with some inaccuracies due to the active nature of the multiplexer circuitry compared to passive de-embedding. Isolated RF probe measurement structures are better suited to determining these aspects of the devices. Certain assumptions can be made about the relationships between DC behavior and these characteristics to significantly reduce the the quantity of devices required for such measurements compared to the DC measurements.}


\gn{A full-fledged cryogenic PDK will need to include additional device types and effects to those discussed here, but the same measurement systems and procedures (multiplexed DC measurement structures and dedicated RF probe structures) are applicable. Chips may be designed to characterize other device types or other types of layout and geometry effects. For example, using farms of multiplexed transistors, parameters related to local layout effects (LLE) may be characterized to include these effects in the PDK. Additionally, passive components and metallization must be characterized. The properties of foundry-supported resistor, inductor, and capacitor structures are subject to significant changes at cryogenic temperatures. Similarly, the characteristics of the metal layers, vias, and contact structures in any technology are likely to change significantly. The characterization of these passive devices are generally simpler than FETs, with a few exceptions; certain materials in standard CMOS processes can become superconducting at cryogenic temperatures\cite{Thomas_2022,noah2023cmos}, meaning that custom compact model equations for these structures would need to be developed to simulate behavior at those conditions.} All of these investigations will be  greatly accelerated by the test system described above.

\section{Conclusion}

\jon{We have presented a measurement scheme for the bulk DC characterization of foundry-fabricated CMOS transistor farms at cryogenic temperatures. This enables a detailed study of device behavior across design variants (e.g. gate length) combined with statistical variations in operating parameters among nominally identical devices. These measurements form the foundation upon which a cryogenic PDK can be built, capable of taking the cryogenic IC design process from  experimental guess-and-check research to a mature workflow capable of optimizing circuits for high performance. The system and measurement scheme presented here provide the advantages of relatively low cost and efficient sample preparation effort compared to existing solutions, making cryogenic PDK development more accessible. The requirements and considerations for preparing such a system have been explored in detail, and measurements conducted on this system have been used for proof-of-concept demonstrations of cryogenic I-V curve-fitting using industry-standard models and cryogenic variability extraction on an intra- and inter-die scale.}

\section{Acknowledgements}
We acknowledge Riyaz Ahamed and Charan Kocherlakota of Quantum Motion for IC layout contributions and James Kirkman of Quantum Motion for PCB design contributions. We acknowledge support from the NPL Quantum Technologies program and the ISCF project Altnaharra (Innovate UK ID 10006186). A. R. acknowledges support from a UKRI Future Leaders Fellowship (MR/T041110/1).

\section{Author Contributions}
J. E. acquired the data. J. E., G. M. N., and D. D. analyzed the data. A. G.-S. designed the ICs. G. M. N. and A. R. packaged the ICs. J. E. and J. D. F. designed the dipping probe. J. E., J. D. F. and G. M. N. developed the measurement software. D. D. extracted model card parameters. A. R., J. D. F., and A. G.-S. supervised the work. G. M. N., A. R., J. D. F., and A. G.-S. conceived the experiments. All authors contributed to the writing of this manuscript.

\bibliographystyle{IEEEtran}
\bibliography{refs}

\newpage
\vspace{11pt}

\vspace{-33pt}
\begin{IEEEbiography}[{\includegraphics[width=1in,height=1.25in,clip,keepaspectratio]{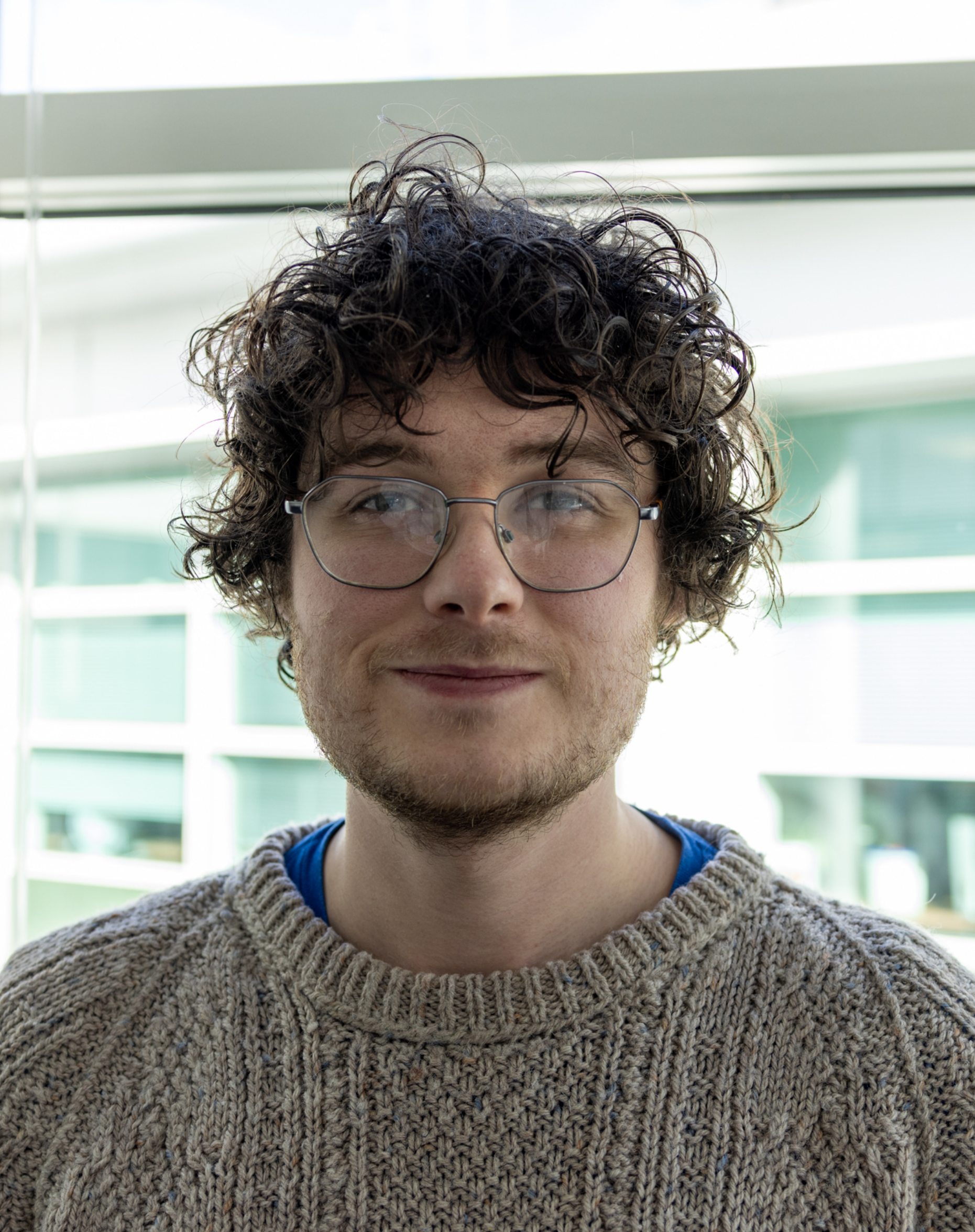}}]
{Jonathan Eastoe} Jonathan received an integrated masters in chemistry and molecular physics from the university of Nottingham in 2018 , where he focused on the development of automated semiconductor test systems. Prior to joining NPL in 2021, Jon worked as part of a product development team designing novel lab instruments for life science applications. Jonathan's research interests include cryogenic CMOS and fA level current calibration. 
\end{IEEEbiography}

\vspace{-33pt}
\begin{IEEEbiography}[{\includegraphics[width=1in,height=1.25in,clip,keepaspectratio]{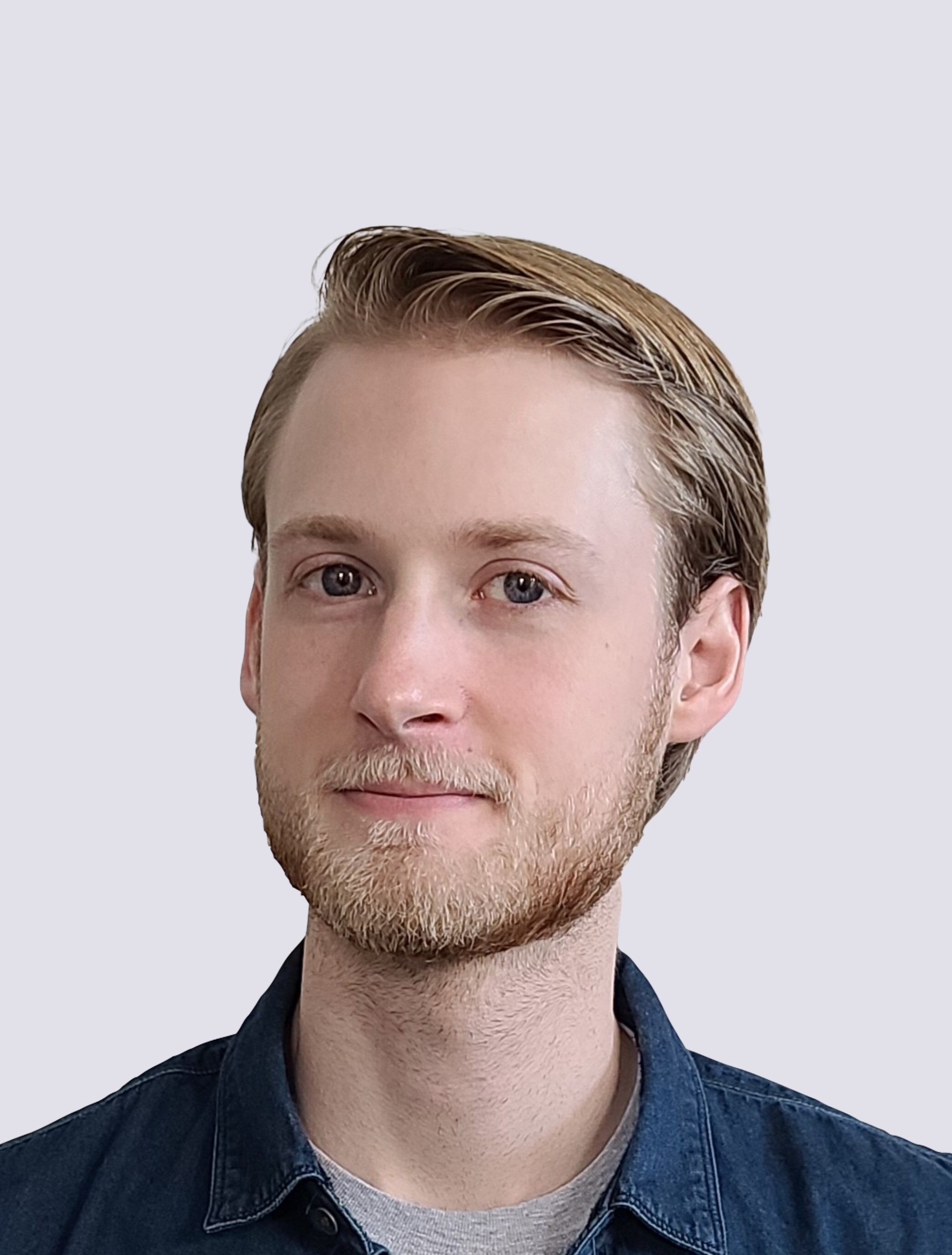}}]{Grayson M. Noah} received the B. S. degree in 2018 and M. S. degree in 2019 from the Georgia Institute of Technology, Atlanta, GA, USA, both in electrical and computer engineering. From 2019 to 2021 he was a Product Engineer at Texas Instruments, Dallas, TX, USA, working on automated test of processors and standardization of latch-up test procedures. Currently leading the Quantum Integration and Validation team at Quantum Motion, his work focuses on measurement, modeling, and thermometry in cryo-CMOS integrated circuits and quantum dot devices.
\end{IEEEbiography}

\vspace{-33pt}
\begin{IEEEbiography}[{\includegraphics[width=1in,height=1.25in,clip,keepaspectratio]{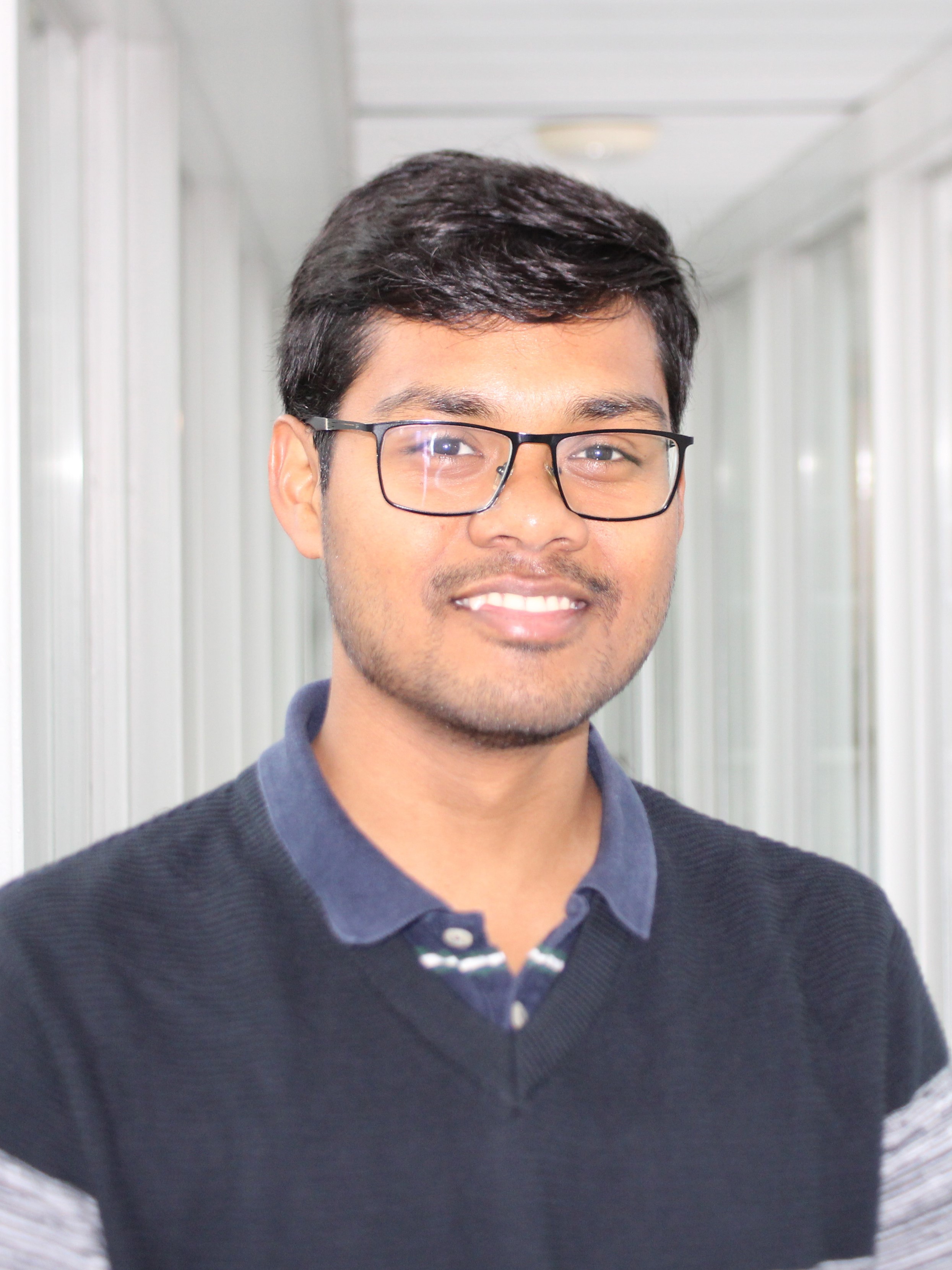}}]{Debargha Dutta} received an MSc. in Microelectronics in 2023 from Delft University of Technology, where he worked on cryogenic RF characterization and modeling of CMOS transistors. His research interest lies around cryogenic integrated circuits for quantum computing applications. His current work at Quantum Motion is in cryo-CMOS characterization and model development.\end{IEEEbiography}

\vspace{-33pt}
\begin{IEEEbiography}[{\includegraphics[width=1in,height=1.25in,clip,keepaspectratio]{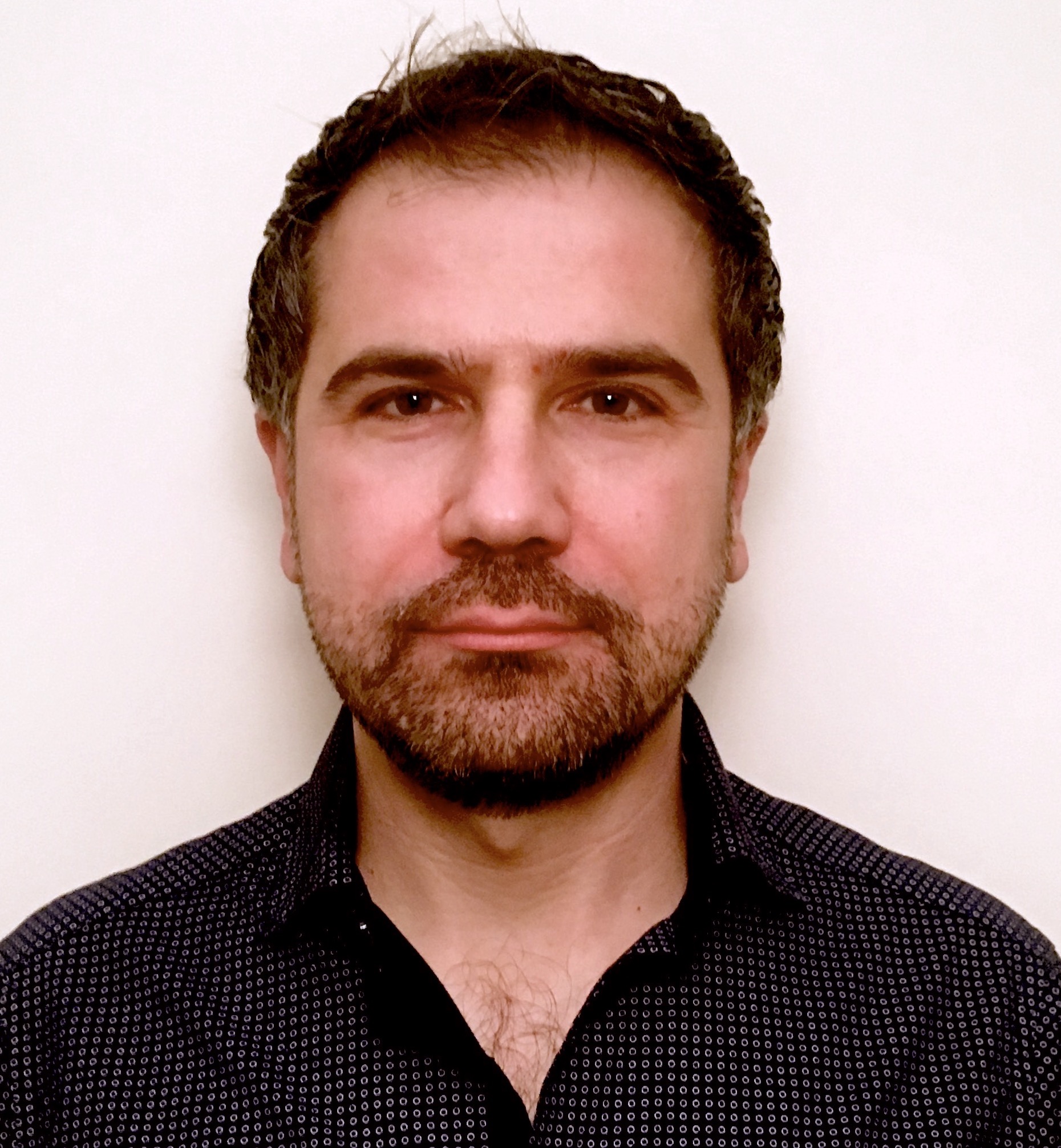}}]{Alessandro Rossi} jointly holds a UKRI Future Leaders Fellowship at the University of Strathclyde (Glasgow, UK) and a Measurement Fellowship at the National Physical Laboratory (London, UK). His research interests range from quantum computing to quantum electrical metrology in semiconductor systems. Alessandro obtained a PhD in Physics from the University of Cambridge (UK) and a MSc summa cum laude in Electronic Engineering from the University of Naples (Italy).\end{IEEEbiography}

\vspace{-33pt}
\begin{IEEEbiography}[{\includegraphics[width=1in,height=1.25in,clip,keepaspectratio]{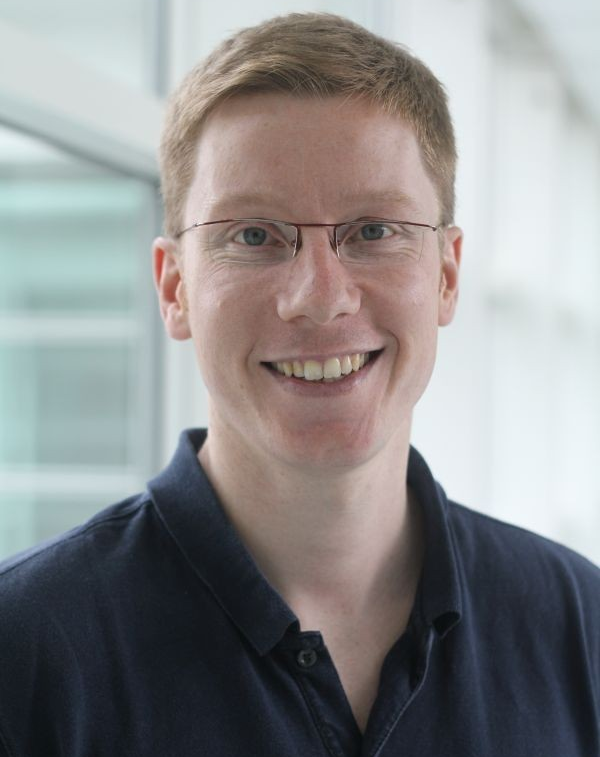}}]{Jonathan D. Fletcher} Jonathan Fletcher was awarded a PhD from the University of Bristol in 2005 for work on unconventional superconductors. He has been a scientist the UK's national measurement institute, the National Physical Laboratory, since 2009. His work focusses on single electron devices for precision electrical metrology and transport of high energy single-electrons in quantum hall edge channels using noise and tomographic techniques.\end{IEEEbiography}

\vspace{-33pt}
\begin{IEEEbiography}[{\includegraphics[width=1in,height=1.25in,clip,keepaspectratio]{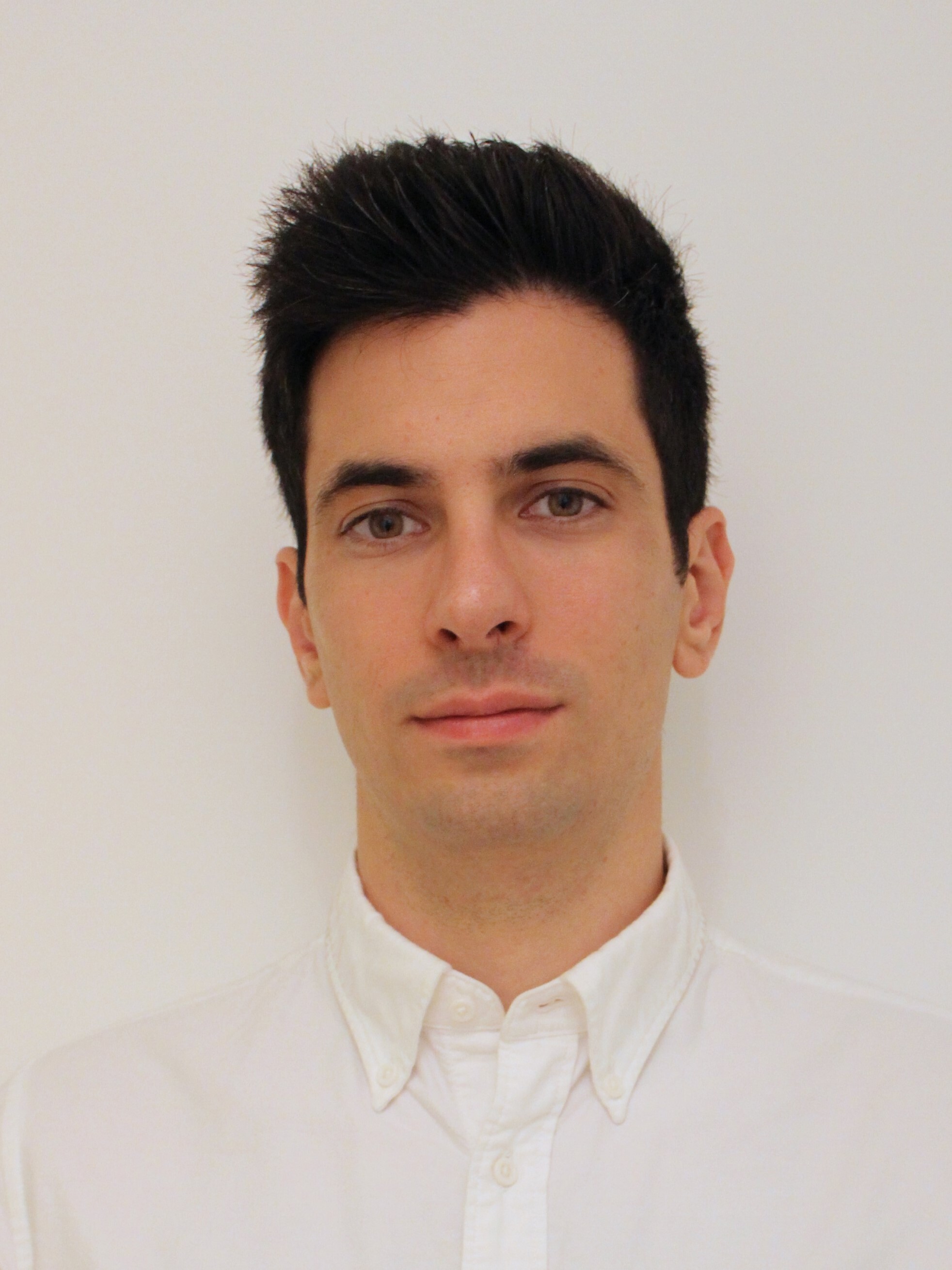}}]{Alberto Gomez-Saiz} holds an MSc. in IC Design from Imperial College London and an MSc. in Quantum Technologies from University College London. He has over 10 years of experience in IC design and leads the IC team at Quantum Motion. During his career he has contributed to the development of the first commercial NFC+Bluetooth SoC (CSR) and the first commercial NB-IoT SoC (Huawei). His research interests lie in cryo-CMOS design for controlling quantum processors and circuit topologies that combine conventional and quantum components.\end{IEEEbiography}

\vfill


\end{document}